\documentclass[conference]{IEEEtran}
\IEEEoverridecommandlockouts

\usepackage{fancyhdr}
\usepackage[numbers, sort]{natbib}
\usepackage{amsmath,amssymb,amsfonts}
\usepackage{algorithmic}
\usepackage{caption}
\usepackage{stfloats} 
\usepackage{multirow}
\usepackage{fancybox}
\usepackage{graphicx}
\usepackage{lipsum}
\usepackage{xspace}
\usepackage{xcolor}
\usepackage{comment}
\usepackage{balance}
\usepackage{enumitem}
\usepackage{hyperref}
\usepackage{cleveref}
\usepackage[labelfont={bf}, font={small}]{caption}

\usepackage{subcaption}
\usepackage{listings}
\usepackage{color}

\definecolor{pblue}{rgb}{0.13,0.13,1}
\definecolor{pgreen}{rgb}{0,0.5,0}
\definecolor{pred}{rgb}{0.9,0,0}
\definecolor{pgrey}{rgb}{0.46,0.45,0.48}

\begin{document}

\title{Extracting Label-specific Key Input Features for Neural Code Intelligence Models}

\author{
    \IEEEauthorblockN{Md Rafiqul Islam Rabin}
    \IEEEauthorblockA{
        University of Houston\\
        Houston, Texas, United States\\
        \textit{mrabin@central.uh.edu}
    }
}

\maketitle
\newcommand{\Part}[1]{\noindent\textbf{#1}}

\cornersize{.2}
\newcounter{observation}
\newcommand{\observation}[1]{\refstepcounter{observation}
        \begin{center}
        \vspace{2pt}
        \Ovalbox{
            \begin{minipage}{0.9\columnwidth}
                \textbf{Observation \arabic{observation}:} #1
            \end{minipage}
        }
        \vspace{2pt}
        \end{center}
}

\newcommand{\ctv}{\textsc{Code2Vec}\xspace}
\newcommand{\cts}{\textsc{Code2Seq}\xspace}

\newcommand{\JS}{\textsc{Java-Small}\xspace}
\newcommand{\JM}{\textsc{Java-Med}\xspace}
\newcommand{\JL}{\textsc{Java-Large}\xspace}
\newcommand{\JTT}{\textsc{Java-Top10}\xspace}

\newcommand{\mnp}{\textsc{MethodName}\xspace}

\newcommand{\perses}{\textsc{Perses}\xspace}
\newcommand{\ddd}{Delta-Debugging\xspace}
\newcommand{\dd}{\textsc{DD}\xspace}

\newcommand{\typeF}{\textsc{Frequent}\xspace}
\newcommand{\typeR}{\textsc{Rare}\xspace}
\newcommand{\typeS}{\textsc{Small}\xspace}
\newcommand{\typeL}{\textsc{Large}\xspace}

\newcommand{\ci}{code intelligence\xspace}
\newcommand{\Ci}{Code intelligence\xspace}
\newcommand{\CI}{Code Intelligence\xspace}

\begin{abstract}
The code intelligence (CI) models are often black-box and do not offer any insights on the input features that they learn for making correct predictions. This opacity may lead to distrust in their prediction and hamper their wider adoption in safety-critical applications. In recent, the program reduction technique is widely being used to identify key input features in order to explain the prediction of CI models. The approach removes irrelevant parts from an input program and keeps the minimal snippets that a CI model needs to maintain its prediction. However, the state-of-the-art approaches mainly use a syntax-unaware program reduction technique that does not follow the syntax of programs, which adds significant overhead to the reduction of input programs and explainability of models.

In this paper, we apply a syntax-guided program reduction technique that follows the syntax of input programs during reduction. Our experiments on multiple models across different types of input programs show that the syntax-guided program reduction technique significantly outperforms the syntax-unaware program reduction technique in reducing the size of input programs. Extracting key input features from reduced programs reveals that the syntax-guided reduced programs contain more label-specific key input features and are more vulnerable to adversarial transformation when renaming the key tokens in programs. These label-specific key input features may help to understand the reasoning of models' prediction from different perspectives and increase the trustworthiness to correct classification given by CI models.
\end{abstract}


\section{Introduction}
\label{sec:introduction}

Deep neural networks have exaggerated their capability to reduce the burden of feature engineering in complex domains, including code intelligence \cite{allamanis2018survey, sharma2021survey}. While the performance of neural models for intelligent code analysis continues to improve, our understanding of what relevant features they learn for correct classification is largely unknown. Therefore, in this paper, we aim to extract key input features that code intelligence models learn for the target label.

A neural code intelligence (CI) model is a deep neural network that takes a program as input and predicts certain properties of that program as output, for example, predicting method name \cite{allamanis2015suggesting}, variable name \cite{allamanis2018ggnn}, or type \cite{hellendoorn2018type} from a program body.
Recent studies have shown that state-of-the-art CI models do not always generalizable to other experiments \cite{kang2019generalizability, rabin2020evaluation}, heavily rely on specific tokens \cite{compton2020obfuscation, rabin2021sivand, suneja2021probing} or structures \cite{rabin2021code2snapshot}, can learn noisy data points \cite{allamanis2019duplication, rabin2021memorization}, and are often vulnerable to semantic-preserving adversarial examples \cite{yefet2020adversarial, rabin2021generalizability}.
Therefore, it is important to know what input features those CI models learn for making correct predictions. The lack of understanding would hinder the trustworthiness to correct classification given by CI models.
Hence, researchers are interested to extract relevant input features that CI models learn for the target label. Such transparency about learned input features is key for wider adoption and application in critical settings such as vulnerability detection or auto-fix suggestion.

Models usually represent an input program as continuous distributed vectors that are computed after training on a large volume of programs. From that, understanding what input features a black-box model has learned is very challenging. For example, code2vec model \cite{alon2019code2vec} learns to represent an input program as a single fixed-length high dimensional embeddings, however, the meaning or characteristics of each dimension are not defined.
An attention-based approach can be used to enhance important code elements in a program. For example, \citet{bui2019autofocus} identify relevant code elements by perturbing statements of the program and combining corresponded attention and confidence scores. However, the attention-based approach poorly correlates with key elements and suffers from a lack of explainability. Recent studies \cite{suneja2021probing, rabin2021sivand, wang2021demystifying} show that the reduction-based approach can extract relevant input features in programs with offering a better explainability.

Several works have already been conducted by researchers for finding relevant input features in models' inference. \citet{allamanis2015suggesting} use a set of hard-coded features from source code and show that extracting relevant features is essential for learning effective code context. \citet{rabin2020demystifying} attempt to find key input features of a label by manually inspecting some input programs of that label. However, the manual inspection cannot be applied to a large dataset due to the vast number of target labels. \citet{suneja2021probing} and \citet{rabin2021sivand} apply a syntax-unaware program reduction technique, \ddd \cite{zeller2002dd}, to reduce the size of input programs in order to find the minimal snippet that a model needs to maintain its prediction. However, this approach creates a large number of invalid and unnatural programs as it does not follow the syntax of programs during the reduction, which adds significant overhead to the explainability of models. While state-of-the-art approaches use a manual inspection or syntax-unaware program reduction technique, we focus on applying the syntax-guided program reduction technique. In particular, we adopt \perses \cite{sun2018perses}, a syntax-guided program reduction technique, to reduce the size of an input program.

In this paper, we apply a syntax-guided reduction technique, rather than syntax-unaware reduction technique, to remove irrelevant parts from an input program and keep the minimal snippet that the CI model needs to maintain its prediction.
Given a model and an input program, our approach adopts \perses \cite{sun2018perses}, a syntax-guided reduction technique, to reduce the size of an input program. The approach continues reducing the input program as long as the model maintains the same prediction on the reduced program as on the original program.
The main insight is that, by reducing some input programs of a label, we may better extract key input features of that target label.
As the syntax-guided technique follows the syntax of input programs, it will always generate valid input programs. Therefore, the approach is more likely to reach the minimal snippet in a smaller number of reduction steps, which will decrease the total reduction time. Moreover, following a syntax-guided technique, the approach can reveal more realistic key input features for the target label.
However, for supporting a programming language data, the syntax-guided technique needs to leverage knowledge about program syntax for avoiding generating syntactically invalid programs.

An experiment with two CI models and four types of input programs reveals that the syntax-guided \perses performs very well compared to the syntax-unaware \ddd. While \perses can generate $100\%$ valid programs, \ddd generates around $10\%$ valid programs only. On average, \perses removes $20\%$ more tokens, takes $70\%$ fewer reduction steps, and spends $2x$ less reduction time than \ddd for reducing an input program.
Furthermore, our results show that we can find label-specific key input features by reducing input programs using \perses, which can provide additional explanation for a prediction and highlight the importance of key input features in programs by triggering $10\%$ more misprediction with $50\%$ fewer adversarial examples.

\noindent\textbf{Contributions.}
This paper makes the following contributions.
\begin{itemize}
    \item We apply a syntax-guided program reduction technique for reducing an input program while preserving the same prediction of the CI model.
    \item We provide a systematic comparison between the syntax-guided program reduction and the syntax-unaware program reduction techniques.
    \item Our results suggest that the syntax-guided program reduction technique significantly outperforms the syntax-unaware program reduction technique.
    \item We highlight key input features that CI models learn for the target label using syntax-guided program reduction.
    \item We show that different program reduction techniques may provide additional explanations for a specific prediction.
\end{itemize}

\section{Related Work}
\label{sec:related}

There has been some work in the area of code intelligence that focuses on the understanding of what relevant features a black-box model learns for correct predictions.
While some work \cite{compton2020obfuscation, rabin2019tnpa, kang2019generalizability, yefet2020adversarial, rabin2021generalizability, suneja2021probing, rabin2021sivand} studies the reliance of models on specific features, many works \cite{allamanis2015suggesting, bui2019autofocus, rabin2020demystifying, suneja2021probing, rabin2021sivand, wang2021demystifying} focus on finding relevant features for explaining models' prediction.

\subsection{Learning Representation of Source Code}
An input program is usually represented as vector embeddings for processing and analyzing by neural models. \citet{allamanis2014learning} introduced a framework that processed token sequences and abstract syntax trees of code to represent the raw programs. \citet{alon2019code2vec} proposed an attention-based neural model that uses a bag of path-context from abstract syntax tree for representing any arbitrary code snippets. \citet{allamanis2018ggnn} constructed data and control flow graphs from programs to encode a code snippet. \citet{hellendoorn2018type} proposed an RNN-based model using sequence-to-sequence type annotations for type suggestion. 
There are some surveys on the taxonomy of models that exploit the source code analysis \cite{allamanis2018survey,sharma2021survey}. \citet{chen2019embeddings} also provide a survey that includes the usage of code embeddings based on different granularities of programs. However, these models are often black-box and do not provide any insight on the meaning or characteristic of learned embeddings. What features or patterns these embeddings represent are largely unknown. In this work, we extract key input features that a model learns for predicting a target label as an explanation of learned embeddings.

\subsection{Reliance on Specific Features}
Models often learn irrelevant features, simple shortcuts, or even noise for achieving target performance.
\citet{compton2020obfuscation} show that the code2vec embeddings highly rely on variable names and cannot embed an entire class rather than an individual method. They investigate the effect of obfuscation on improving code2vec embeddings that better preserves code semantics. They retrain the code2vec model with obfuscated variables to forcing it on the structure of code rather than variable names and aggregate the embeddings of all methods from a class.
Following the generalizability of word embeddings, \citet{kang2019generalizability} assess the generalizability of code embeddings in various software engineering tasks and demonstrate that the learned embeddings by code2vec do not always generalizable to other tasks beyond the example task it has been trained for.
\citet{rabin2021generalizability} and \citet{yefet2020adversarial} demonstrate that the models of code often suffer from a lack of robustness and be vulnerable to adversarial examples. They mainly introduce small perturbations in code for generating adversarial examples that do not change any semantics and find that the simple renaming, adding or removing tokens changes model's predictions.
\citet{suneja2021probing} uncover the model’s reliance on incorrect signals by checking whether the vulnerability in the original code is missing in the reduced minimal snippet. They find that model captures noises instead of actual signals from the dataset for achieving high predictions. \citet{rabin2021sivand} demonstrates that models often use just a few simple syntactic shortcuts for making prediction. \citet{rabin2021memorization} also show that models can fit noisy training data with excessive parameter capacity. 
As models often learn noise or irrelevant features for achieving high prediction performance, the lack of understanding of what input features models learn would hinder the trustworthiness to correct classification. Such opacity is substantially more problematic in critical applications such as vulnerability detection or auto-fix suggestion. In this work, we extract key input features for CI models in order to provide better transparency and explaining the predictions.

\subsection{Extracting Relevant Input Features}
Several kinds of research have been done in finding relevant input features for models of source code.
\citet{allamanis2015suggesting} exhibit that extracting relevant features is essential for learning effective code context. They use a set of hard-coded features from source code that integrate non-local information beyond local information and train a neural probabilistic language model for automatically suggesting names. However, extracting hard-coded features from source code may not be available for arbitrary code snippets and in dynamically typed languages.
\citet{bui2019autofocus} propose a code perturbation approach for interpreting attention-based models of source code. It measures the importance of a statement in code by deleting it from the original code and analyzing the effect on predicted outputs. However, the attention-based approach often poorly correlates with key elements and suffers from a lack of explainability.
\citet{rabin2020demystifying} attempt to find key input features of a label by manually inspecting some input programs of that label. They extract handcrafted features for each label and train simple binary SVM classification models that achieves highly comparable results to the higher dimensional code2vec embeddings for the method naming task. However, the manual inspection cannot be applied to a large dataset.
\citet{wang2021demystifying} propose a mutate-reduce approach to find key features in the code summarization models.
\citet{suneja2021probing} and \citet{rabin2021sivand} apply a syntax-unaware program reduction technique, Delta Debugging \cite{zeller2002dd}, to find minimal snippet which a model needs to maintain its prediction. By removing irrelevant parts to a prediction from the input programs, the authors aim to better understand important features in the model inference. 
However, the syntax-unaware approach creates a large number of invalid and unnatural programs during the reduction as it does not follow the syntax of programs, thus increases the total steps and time of reduction. In this work, we apply a syntax-guided program reduction technique that overcomes the overhead raised by the syntax-unaware technique.

\section{Design and Implementation}
\label{sec:design}

\begin{figure*}
    \centering
    \captionsetup{font=large}
    \includegraphics[width=0.9\textwidth]{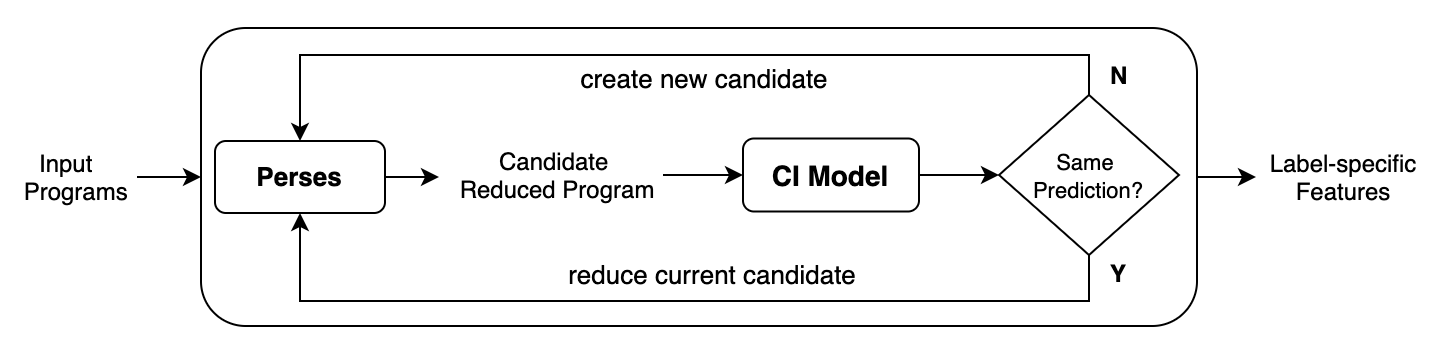}
    \caption{Workflow of our approach.}
    \label{fig:approach}
\end{figure*}

This section describes our approach of extracting input features for code intelligence (CI) models by syntax-guided program reduction. We use \perses \cite{sun2018perses} as the syntax-guided program reduction technique in our study. We first provide an overview of how \perses works and then describe how we adopt it in the workflow of our approach.

\medskip
\Part{\perses.}
\citet{sun2018perses} have proposed the framework for syntax-guided program reduction called \perses. 
Given an input program, the grammar of that programming language, and the output criteria, \perses reduces the input program with respect to the grammar while preserving the output criteria. It mainly follows the below steps.
\begin{itemize}
    \item It first parses the input program into a parse tree by normalizing the definition of grammar. 
    \item Then it traverses the tree and determines whether a tree node is deletable (such as follows the grammar and preserves the output criteria). If yes, it prunes the sub-tree from that node and generates a valid reduced program, else it ignores that node and avoids generating invalid programs. Thus, in each iteration of reduction, it ensures generating syntactically valid program variants that preserves the same output criteria. 
    \item Next, the deletion of one node may enable the deletion of another node. Therefore, \perses is repeatedly applied to the reduced program until no more tree nodes can be removed, which is known as fixpoint mode reduction.
    \item The final reduced program is called 1-tree-minimal, and any further attempts to reduce the program would generate an invalid program or change the output criteria.
\end{itemize}
We integrate the \perses as a black-box framework in our approach for extracting input features of CI models.

\medskip
\Part{Workflow.}
Figure~\ref{fig:approach} depicts a high-level view of the workflow in our proposed methodology.
Given a set of input programs, our approach reduces each input program using \perses while preserving the same prediction by the CI model. 
The approach removes irrelevant parts from an input program and keeps the minimal code snippet that the CI model needs to maintain its prediction. 
The main insight is that, by reducing some input programs of a target label, we can identify key input features of the CI model for that target label.
Our approach follows the below steps.
\begin{itemize}
    \item Given an input program $P$ and a CI model $M$, our approach first record the prediction $y$ (i.e. predicted method name) given by the CI model $M$ on the input program $P$, such as $y = M(P)$.
    \item Using \perses, we then generate a candidate reduced program $R'$ by removing some nodes from the tree of the input program $P$, such as $R' = \perses(P)$.
    \item If the candidate reduced program $R'$ does not hold the same prediction $y$ by the CI model $M$ (i.e. $y \neq M(R')$), we reject this candidate program and create another candidate program by removing some other nodes from the tree of the input program.
    \item If the candidate reduced program $R'$ preserves the same prediction $y$ by the CI model $M$ (i.e. $y = M(R')$), we continue reduction and iteratively search for the final reduced program $R$ that produces the same prediction $y$.
    \item The final reduced program is 1-tree-minimal, which contains the key input features that the CI model must need for making the correct prediction $y$.
\end{itemize}
After reducing a set of input programs of a target label, we extract the node type and token value from the abstract syntax tree (AST) of each reduced program.
Every extracted element from reduced programs is considered as a \textit{candidate} input feature.
The most common elements are identified as label-specific \textit{key} input features and other uncommon elements are identified as input-specific \textit{sparse} features.

\medskip
\Part{Implementation.}
Our approach is model-agnostic and can be applied for various tasks and programming datasets.
In this paper, for experimentation of our approach, we study two well-known code intelligence models (\ctv and \cts), a popular code intelligence task (\mnp) and one commonly used programming language dataset (\JL) with different types of input programs. This section outlines all of these.

\subsubsection{Task}
\begin{figure}[t]
    \centering
    \includegraphics[width=0.95\columnwidth]{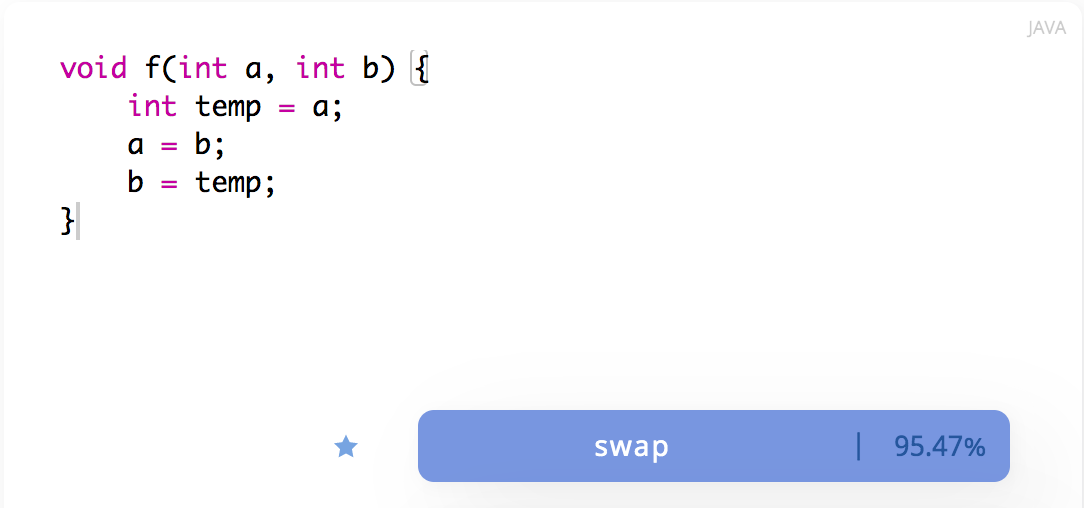}
    \vspace{1mm}
    \caption{An example of \mnp task by \ctv\cite{alon2019code2vec}.}
    \label{fig:example}
\end{figure}

We use the method name prediction (\mnp \cite{allamanis2015suggesting}) task in this study. 
This task is commonly used by researchers in the code intelligence domain for various applications such as code summarization \cite{allamanis2015suggesting,allamanis2016summarization}, representation learning \cite{alon2019code2vec,alon2019code2seq}, neural testing \cite{kang2019generalizability,yefet2020adversarial,rabin2021generalizability}, feature extraction \cite{rabin2021sivand,wang2021demystifying}, and so on \cite{allamanis2018survey,sharma2021survey}.
In the \mnp task, a model attempts to predict the name of a method from its body. \Cref{fig:example} shows an example of \mnp task, where given the following code snippet: ``\texttt{void f(int a, int b) \{int temp = a; a = b; b = temp;\}}'', the \ctv model correctly predicts the method's name as ``\texttt{swap}''.

\subsubsection{Models}
We use the \ctv~\cite{alon2019code2vec} and \cts~\cite{alon2019code2seq} \ci models for \mnp task. Both models use paths from abstract syntax trees (AST) to encode a program. Given a sample expression ``\texttt{a = b;}'', an example of path context in AST is ``\texttt{a, <NameExpr ↑ AssignExpr ↓ IntegerLiteralExpr>, b}''.
\begin{itemize}
    \item \ctv. This model extracts a bag of path-context from the AST of the program where each path-context includes a pair of terminal nodes and the corresponding path between them. The model learns embeddings of these path-contexts during training and uses an attention mechanism to aggregate multiple path-contexts to a single code vector. The code vector is used as a representation of the program for making a prediction.
    \item \cts. This model also extracts a bag of path-context from the AST of the program but it sub-tokenized each path-context. The \cts model uses a bi-directional LSTM to encode paths node-by-node, and another LSTM to decode a target sequence one-by-one.
\end{itemize}

\subsubsection{Dataset}
For the \mnp task, we use the \JL dataset \cite{alon2019code2seq}. This dataset contains a total of $9,500$ Java projects from GitHub, where $9,000$ projects are for the training set, $200$ projects for the validation set, and $300$ projects for the test set. Using training set and validation set, we train both the \ctv and \cts models.

\subsubsection{Input Types}
The dataset from GitHub is often imbalanced and contains different sizes and frequencies of input programs. Therefore, we choose different types of input programs from the \JL test set to evaluate the effectiveness of our approach in terms of reduction and feature extraction.

\begin{itemize}
    \item Frequent Methods: We randomly sample a total of $100$ input programs from the most occurring method names.
    \item Rare Methods: We randomly sample a total of $100$ input programs from the least occurring method names.
    \item Smaller Methods: We randomly sample a total of $100$ input programs that contains less than $10$ lines of code.
    \item Larger Methods: We randomly sample a total of $50$ input programs that has around $100$ lines of code.
\end{itemize}

Moreover, to demonstrate the label-specific key input features, we select correctly predicted input programs from the ten most frequent labels of the \JL test set for feature extraction. Those labels (methods) are: equals, main, setUp, onCreate, toString, run, hashCode, init, execute, and get.

\subsubsection{Syntax-unaware Reduction Technique}
We use the \ddd algorithm as the syntax-unaware program reduction technique in this study. \citet{zeller2002dd} have proposed the \ddd algorithm to reduce the size of an input program. The algorithm iteratively splits an input program into multiple candidate programs by removing parts of the input program. The algorithm then checks if any resulting candidate program preserves the prediction of the model on the original input program. When the algorithm finds a candidate satisfying the property, it uses the candidate as the new base to be reduced further. Otherwise, the algorithm increases the granularity for splitting, until it determines that the input program cannot be reduced further.

\begin{itemize}
    \item \textit{DD-Token}: In the token level approach, \ddd reduces the size of an input program token by token. We mostly use the \textit{DD-Token} as the default baseline for \ddd in this study.
    \item \textit{DD-Char}: In the char level approach, \ddd reduces the size of an input program char by char. We use the \textit{DD-Char} approach to provide an additional explanation in \Cref{sec:results_explanation} and \Cref{code:explanation_main}.
\end{itemize}

\citet{rabin2021sivand} described more detail on how the \ddd technique is adopted in the workflow of reducing input programs for CI models.

\section{Results}
\label{sec:results}

In this section, we present the average result of our experiments on the \ctv and \cts models and the \JL dataset for different input types.

\begin{figure*}
\centering
\captionsetup{font=large}
\begin{subfigure}{.44\textwidth}
  \centering
  \includegraphics[width=\linewidth]{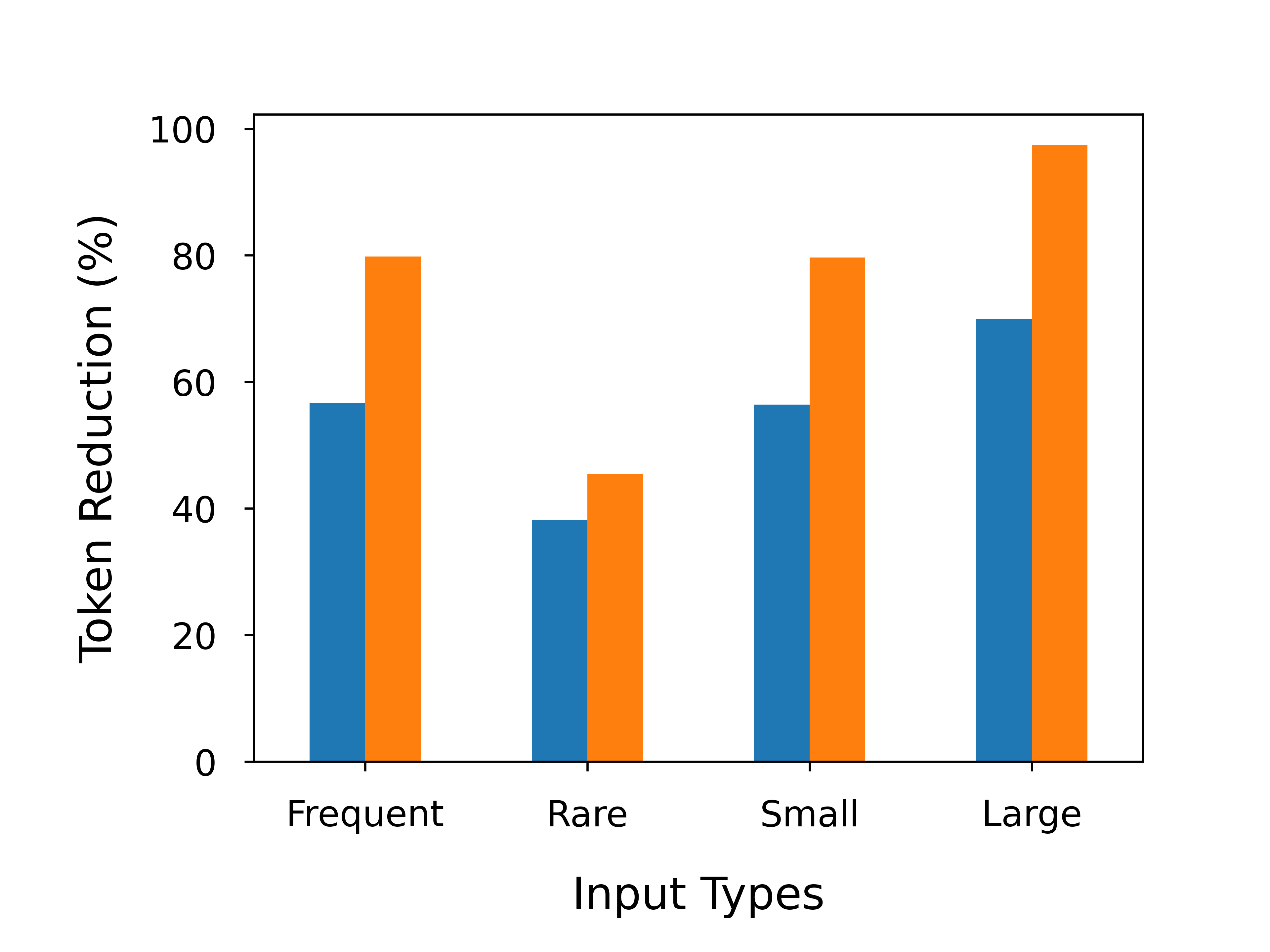}
  \caption{Token Reduction}
  \label{fig:reduction}
\end{subfigure}%
\begin{subfigure}{.44\textwidth}
  \centering
  \includegraphics[width=\linewidth]{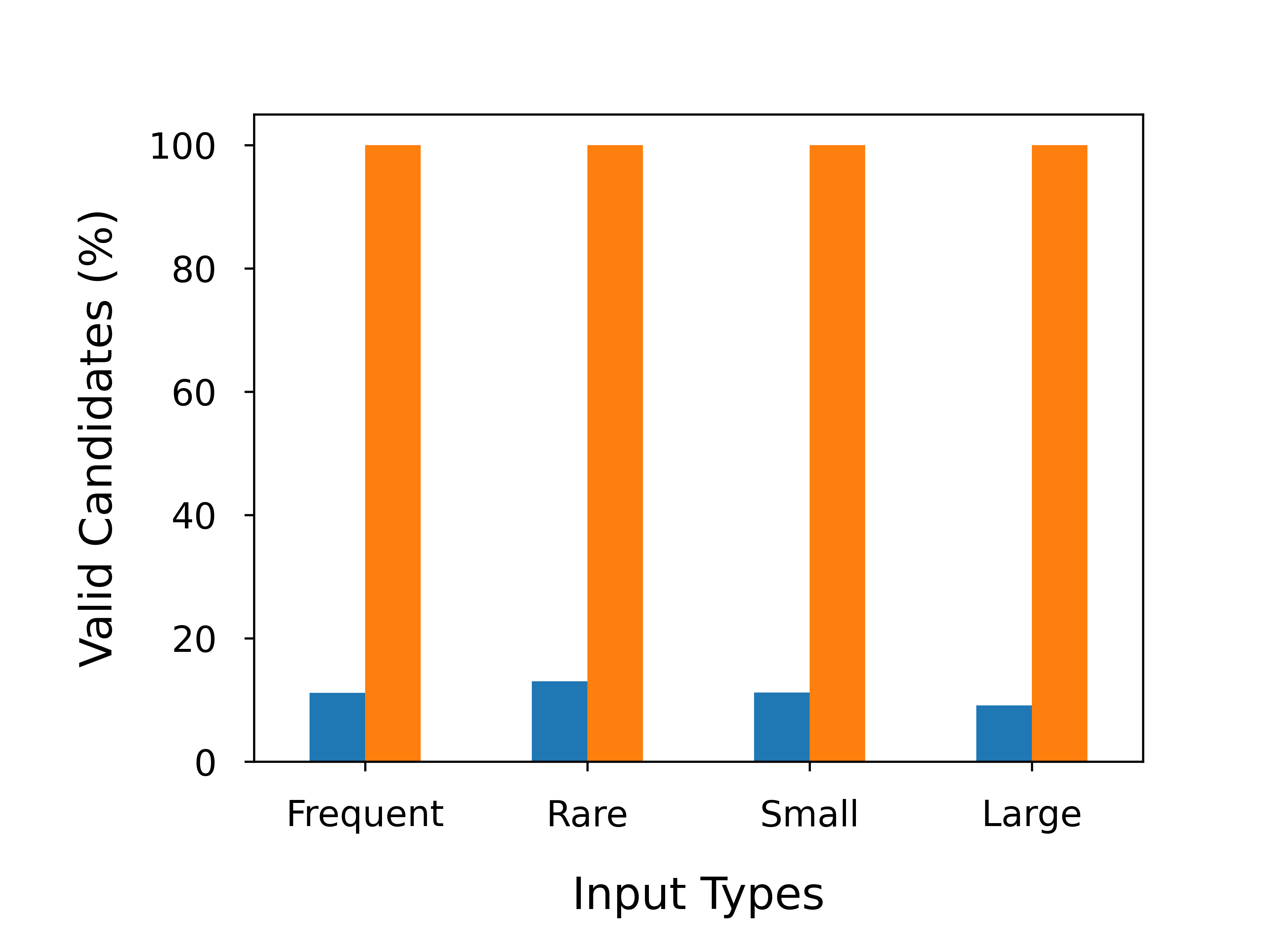}
  \caption{Valid Candidates}
  \label{fig:valid}
\end{subfigure}
\begin{subfigure}{.44\textwidth}
  \centering
  \includegraphics[width=\linewidth]{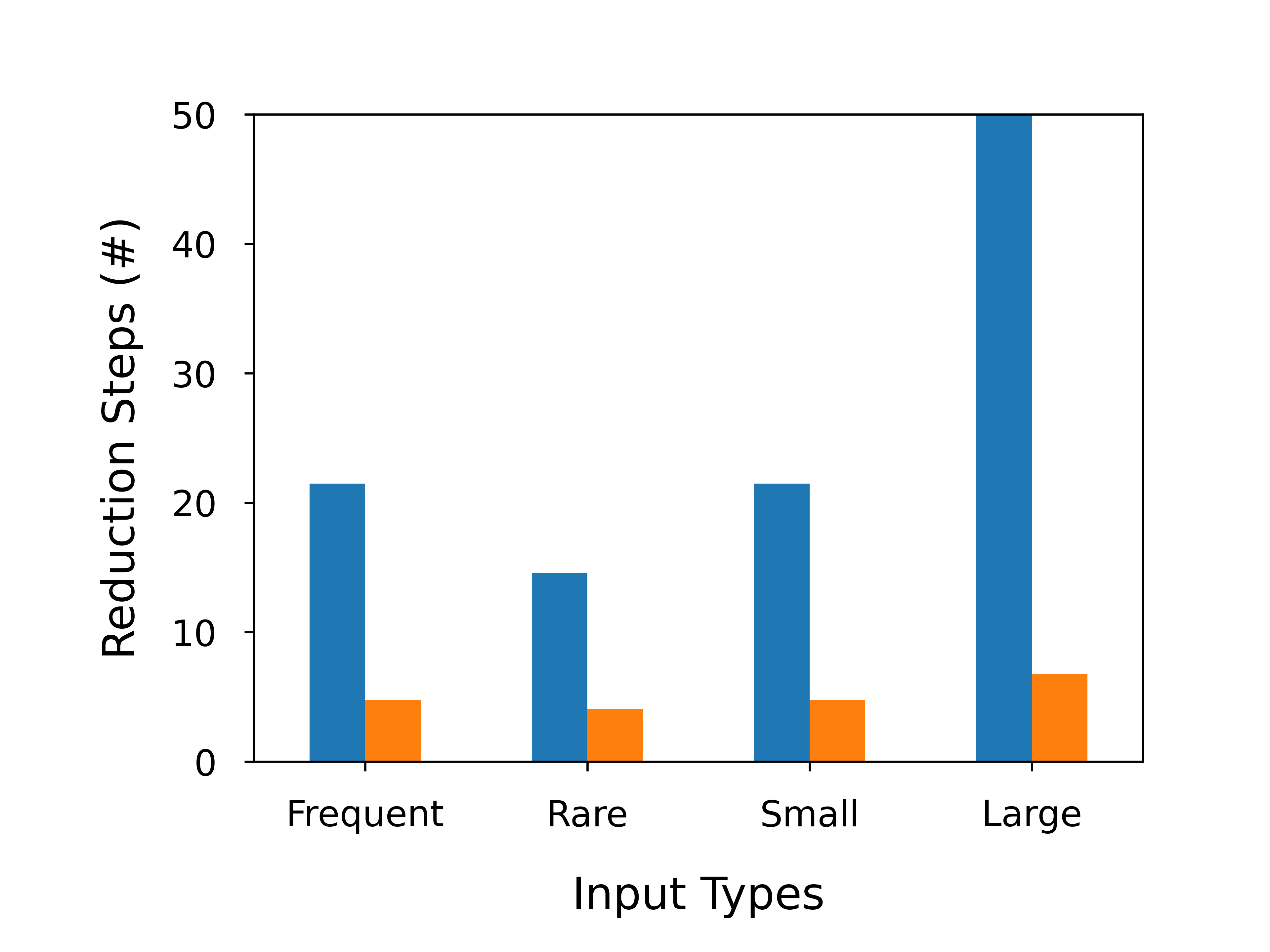}
  \caption{Reduction Steps}
  \label{fig:steps}
\end{subfigure}%
\begin{subfigure}{.44\textwidth}
  \centering
  \includegraphics[width=\linewidth]{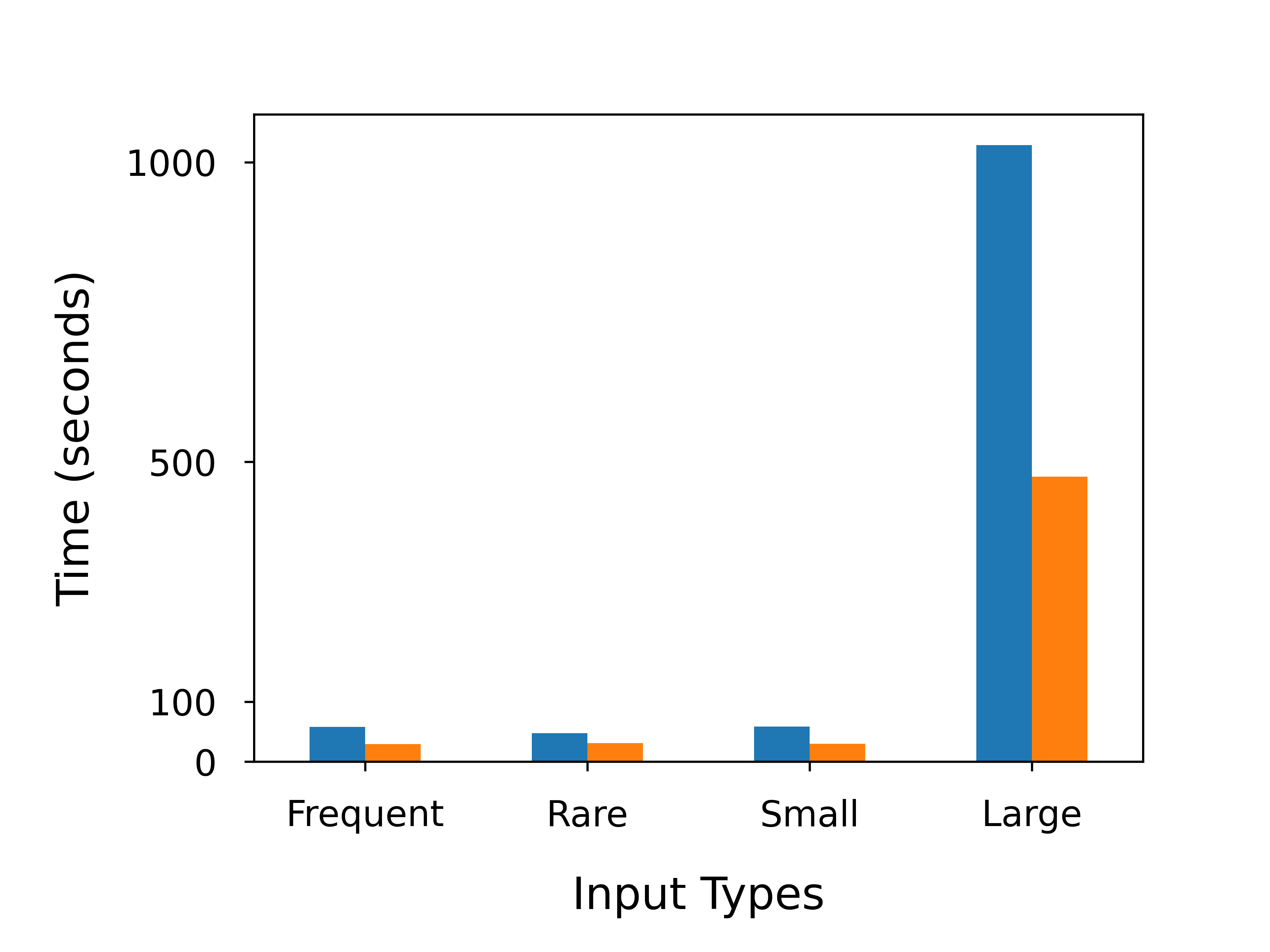}
  \caption{Reduction Time}
  \label{fig:time}
\end{subfigure}
\caption{Comparison between \ddd (\textcolor{blue}{\textbf{blue bar}}) and \perses (\textcolor{orange}{\textbf{orange bar}}).}
\label{fig:comparison}
\end{figure*}

\subsection{Comparative Analysis}
Here, we provide a systematic comparison between the syntax-guided program reduction technique and the syntax-unaware program reduction technique.
In particular, we compare the syntax-guided \perses and the syntax-unaware \ddd in terms of token reduction, valid candidates, reduction steps and reduction time.

\subsubsection{Token Reduction}
\label{sec:results_reduction}
The goal of \perses and \ddd is to remove irrelevant tokens from an input program as much as possible while preserving the same prediction of the CI model. 
\Cref{fig:reduction} shows their such ability in reducing the size of the original input programs for different input types. We can see that, for all input types, \perses reduces more tokens from an input program than \ddd. On average, \perses removes $20\%$ more tokens from an input program than \ddd. The difference is most significant (around $30\%$) in \typeL input types and less significant (around $5\%$) in \typeR input types.
This result suggests that \perses is more powerful than \ddd in reducing the size of an input program.

\subsubsection{Valid Candidates}
\label{sec:results_valid}
In each reduction step, \perses and \ddd create a candidate program after removing some irrelevant tokens from an input program, and continue for further reduction. 
\Cref{fig:valid} shows their effectiveness in generating valid candidate programs during reduction. For all input types, \perses always reduces to a valid candidate program (thus, $100\%$ valid candidates) as it follows the syntax of programs during reduction. However, in most cases, \ddd reduces to an invalid candidate program (only around $10\%$ valid candidates) as it does not follow the syntax of programs. Therefore, after each invalid step, \ddd backtracks to the previous step and generates another candidate program by removing tokens from some other parts of the program, which increases the overhead in total reduction steps and reduction time.

\subsubsection{Reduction Steps}
\label{sec:results_steps}
The reduction is applied repeatedly to an input program until finding the final minimal program, from where no more tokens can be removed. 
From \Cref{fig:steps}, we can see that \perses on average can reach the final minimal program within $5$ reduction steps. However, \ddd makes around $20$ reductions in \typeF-\typeR-\typeS input types and more than $50$ reductions in \typeL input type, to reach the final minimal program. The \ddd reduces an input program by a sequence of tokens where \perses can prune an entire sub-tree from AST. Thus, \perses takes a much lower number of reduction steps than \ddd to reach the final minimal program.

\subsubsection{Reduction Time}
\label{sec:results_time}
We now compare the average time taken by \perses and \ddd for reducing an input program. As \ddd takes excessive invalid steps, \perses is expected to spend less time for program reduction.
\Cref{fig:time} shows that, for all input types, \perses reduces an input program faster than \ddd, specially in \typeL input type. In \typeF-\typeR-\typeS input types, both \perses and \ddd spend less than $2$ minutes to reduce an input program and comparatively \perses takes $30$ seconds less time than \ddd. In \typeL input types, \ddd spends around $17$ minutes for the reduction of a large program but \perses takes only $8$ minutes, which is around $50\%$ less than \ddd.

\observation{\perses allows more token removal than \ddd and always creates valid candidate programs. Compared to \ddd,  \perses is more likely to reach the final minimal program in a smaller number of reduction steps, which decreases the total reduction time.}

            
            

\begin{table} 
    \begin{center}
        \def\arraystretch{1.2}
        \caption{Summary of key input features in Top-10 methods.}
        \label{table:feature_summary}
        \resizebox{0.9\columnwidth}{!}{%
        \begin{tabular}{|c|c|c|c|c|c|}
            \hline
            \multirow{2}{*}{\textbf{Method}} & \multirow{2}{*}{\textbf{Model}} & \multicolumn{2}{c|}{\textbf{\ddd (DD)}} & \multicolumn{2}{c|}{\textbf{\perses}} \\ \cline{3-6}
             & & \textbf{Candidate} & \textbf{Key} & \textbf{Candidate} & \textbf{Key} \\ \hline
            \hline
            
            \multirow{3}{*}{equals}
            & \ctv & 30  &  8  &  12  &  5 \\ \cline{2-6}
            & \cts & 28  &  8  &  10  &  5 \\ \cline{2-6}
            &(both)& 30  &  8  &  12  &  6 \\ \hline
            
            \multirow{3}{*}{main}
            & \ctv & 27  &  5  &  21  &  5 \\ \cline{2-6}
            & \cts & 27  &  5  &   4  &  3 \\ \cline{2-6}
            &(both)& 28  &  5  &  21  &  5 \\ \hline
            
            \multirow{3}{*}{setUp}
            & \ctv & 36  &  4  &  13  &  5 \\ \cline{2-6}
            & \cts & 27  &  4  &  13  &  1 \\ \cline{2-6}
            &(both)& 41  &  4  &  19  &  5 \\ \hline
            
            \multirow{3}{*}{onCreate}
            & \ctv & 31  &  5  &  20  &  4 \\ \cline{2-6}
            & \cts & 24  &  4  &  14  &  3 \\ \cline{2-6}
            &(both)& 34  &  5  &  24  &  4 \\ \hline
            
            \multirow{3}{*}{toString}
            & \ctv & 21  &  4  &  18  &  5 \\ \cline{2-6}
            & \cts & 22  &  3  &  18  &  2 \\ \cline{2-6}
            &(both)& 23  &  4  &  25  &  5 \\ \hline
            
            \multirow{3}{*}{run}
            & \ctv & 30  &  5  &  22  &  5 \\ \cline{2-6}
            & \cts & 31  &  6  &  13  &  2 \\ \cline{2-6}
            &(both)& 36  &  6  &  27  &  5 \\ \hline
            
            \multirow{3}{*}{hashCode}
            & \ctv & 13  &  5  &   6  &  5 \\ \cline{2-6}
            & \cts & 14  &  5  &  12  &  4 \\ \cline{2-6}
            &(both)& 15  &  5  &  13  &  5 \\ \hline
            
            \multirow{3}{*}{init}
            & \ctv & 30  &  3  &  29  &  3 \\ \cline{2-6}
            & \cts & 18  &  1  &  10  &  1 \\ \cline{2-6}
            &(both)& 35  &  3  &  33  &  3 \\ \hline
            
            \multirow{3}{*}{execute}
            & \ctv & 20  &  5  &  14  &  5 \\ \cline{2-6}
            & \cts & 16  &  3  &   7  &  3 \\ \cline{2-6}
            &(both)& 24  &  6  &  14  &  5 \\ \hline
            
            \multirow{3}{*}{get}
            & \ctv & 56  &  6  &  49  &  4 \\ \cline{2-6}
            & \cts & 23  &  0  &  16  &  0 \\ \cline{2-6}
            &(both)& 58  &  6  &  50  &  4 \\ \hline
            \hline
            \textbf{Top-10} 
            &(both)& \textbf{324} & \textbf{52} & \textbf{238} & \textbf{47} \\ \hline
        \end{tabular}%
        }
    \end{center}
\end{table}

\begin{table} 
    \begin{center}
        \def\arraystretch{1.2}
        \caption{Label-specific key input features in Top-10 methods.}
        \label{table:feature_list}
        \resizebox{0.98\columnwidth}{!}{%
        \begin{tabular}{|c|c|c|l| }
            \hline
            \textbf{Method} & \textbf{Model} & \textbf{Reduction} & \textbf{Key Input Features} \\ \hline
            \hline
            
            \multirow{4}{*}{equals}
            & \multirow{2}{*}{\ctv} 
            & DD & if, boolean, Object, o, obj, other, instanceof, Stock \\ \cline{3-4}
            & & \perses & boolean, Object, return, o, obj \\ \cline{2-4}
            & \multirow{2}{*}{\cts} 
            & DD & if, boolean, Object, obj, other, o, instanceof, Stock \\ \cline{3-4}
            & & \perses & boolean, Object, Override, o, obj \\ \hline
            
            \multirow{4}{*}{main}
            &  \multirow{2}{*}{\ctv} 
            & DD & args, void, String, Exception, throws \\ \cline{3-4}
            & & \perses & void, String, args, System, Exception \\ \cline{2-4}
            &  \multirow{2}{*}{\cts} 
            & DD & args, void, String, Exception, throws \\ \cline{3-4}
            & & \perses & void, String, args \\ \hline
            
            \multirow{4}{*}{setUp}
            &  \multirow{2}{*}{\ctv} 
            & DD & void, throws, Exception, setUp \\ \cline{3-4}
            & & \perses & void, throws, Exception, super, setUp \\ \cline{2-4}
            &  \multirow{2}{*}{\cts} 
            & DD & void, throws, Exception, setUp \\ \cline{3-4}
            & & \perses & void \\ \hline
            
            \multirow{4}{*}{onCreate}
            &  \multirow{2}{*}{\ctv} 
            & DD & void, savedInstanceState, Bundle, onCreate, if \\ \cline{3-4}
            & & \perses & void, savedInstanceState, super, onCreate \\ \cline{2-4}
            &  \multirow{2}{*}{\cts} 
            & DD & void, savedInstanceState, onCreate, Bundle \\ \cline{3-4}
            & & \perses & void, super, onCreate \\ \hline

            \multirow{4}{*}{toString}
            &  \multirow{2}{*}{\ctv} 
            & DD & String, if, toString, sb \\ \cline{3-4}
            & & \perses & String, Override, StringBuilder, sb, return \\ \cline{2-4}
            &  \multirow{2}{*}{\cts} 
            & DD & String, toString, if \\ \cline{3-4}
            & & \perses & String, return \\ \hline
            
            \multirow{4}{*}{run}
            &  \multirow{2}{*}{\ctv} 
            & DD & void, try, catch, 0, x \\ \cline{3-4}
            & & \perses & void, Override, try, catch, x \\ \cline{2-4}
            &  \multirow{2}{*}{\cts} 
            & DD & void, try, catch, 0, Override, x \\ \cline{3-4}
            & & \perses & void, Override \\ \hline
            
            \multirow{4}{*}{hashCode}
            &  \multirow{2}{*}{\ctv} 
            & DD & int, hashCode, 0, result, null \\ \cline{3-4}
            & & \perses & int, Override, result, final, prime \\ \cline{2-4}
            &  \multirow{2}{*}{\cts} 
            & DD & int, hashCode, result, 0, null \\ \cline{3-4}
            & & \perses & int, result, Override, prime \\ \hline
            
            \multirow{4}{*}{init}
            &  \multirow{2}{*}{\ctv} 
            & DD & void, throws, ServletException \\ \cline{3-4}
            & & \perses & void, throws, ServletException \\ \cline{2-4}
            &  \multirow{2}{*}{\cts} 
            & DD & void \\ \cline{3-4}
            & & \perses & void \\ \hline
            
            \multirow{4}{*}{execute}
            &  \multirow{2}{*}{\ctv} 
            & DD & void, throws, BuildException, execute, context \\ \cline{3-4}
            & & \perses & void, throws, BuildException, super, execute \\ \cline{2-4}
            &  \multirow{2}{*}{\cts} 
            & DD & void, execute, super \\ \cline{3-4}
            & & \perses & void, super, execute \\ \hline
            
            \multirow{4}{*}{get}
            &  \multirow{2}{*}{\ctv} 
            & DD & if, T, null, return, key, Object \\ \cline{3-4}
            & & \perses & T, throw, key, return \\ \cline{2-4}
            &  \multirow{2}{*}{\cts} 
            & DD & None\\ \cline{3-4}
            & & \perses & None \\ \hline
            
        \end{tabular}%
        }
    \end{center}
\end{table}

\subsection{Label-Specific Key Input Features}
\label{sec:results_features}

Here, we provide the summary of extracted input features that CI models learn for predicting the target method name. In our experiment, we consider all tokens in reduced programs as \textit{candidate} tokens. A label-specific \textit{key} input feature is a candidate token that appears in at least $50\%$ of reduced programs, where other infrequent tokens are input-specific \textit{sparse} features. 
For brevity and page limit, we only show the Top-10 most frequent methods in \Cref{table:feature_summary} and \Cref{table:feature_list}.

From \Cref{table:feature_summary}, considering both \ctv and \cts models, we can see that both \perses and \ddd identify around $50$ tokens, in total, as label-specific key input features in Top-10 methods.
However, \ddd contains a total of $324$ candidate tokens in reduced programs, which is $1.36$x time higher than \perses that contains a total of $238$ candidate tokens.
In some methods, i.e. `equals' and `setUp', the total number of candidate tokens in \ddd reduced programs is almost $2$x time higher than the candidate tokens in \perses reduced programs. This shows that the tokens found from the reduced programs of \ddd are more input-specific while the tokens found from the reduced programs of \perses are more label-specific.

Furthermore, \Cref{table:feature_list} shows the label-specific key input features (sorted by their frequency) extracted by \ddd and \perses from its reduced programs. These label-specific key input features can help to understand the prediction of the CI model for a target label. For example, \ddd and \perses reveal that ``\texttt{void, args, String, Exception}'' are key features for the `main' method. It highlights that a sample input program containing those tokens is more likely to be predicted as the `main' method by CI models.


\observation{\perses reveals more label-specific key input features in its syntax-guided reduced programs, while \ddd contains more input-specific sparse features in its syntax-unaware reduced programs.}


\begin{figure}
\textbf{(a) A sample input program:}
\begin{lstlisting}
public static void f(String[] args) {
    System.setProperty(
        Constants.DUBBO_PROPERTIES_KEY, 
        "conf/dubbo.properties");
    Main.main(args);
}
\end{lstlisting}

\textbf{(b) DD-Char reduced program:}
\begin{lstlisting}
d f(Sg[]r){y(C,"");Main(ar);}
\end{lstlisting}

\textbf{(c) DD-Token reduced program:}
\begin{lstlisting}
void f(String[]args){("");(args);}
\end{lstlisting}

\textbf{(d) \perses reduced program:}
\begin{lstlisting}
void f(String args) { }
\end{lstlisting}

\caption{A sample input program and corresponding reduced programs for different program reduction techniques.}
\label{code:explanation_main}
\end{figure}

\subsection{Multiple Explanation for a Specific Prediction}
\label{sec:results_explanation}

Different program simplification approaches, i.e., \ddd and \perses, provide a different set of key features for a target label by a CI model (\Cref{table:feature_list}). Those different features can help us to find multiple explanations for a specific prediction. 
For instance, the \cts predicts the code snippet in \Cref{code:explanation_main}a as the \texttt{main} method. 
The \ddd with char-based program reduction (DD-Char) gives the minimal program in \Cref{code:explanation_main}b, that \cts can predict as \texttt{main}. We can see the presence of the \texttt{Main} identifier in the method body of \Cref{code:explanation_main}b which is one of the important tokens for the target prediction. On the other hand, the \ddd with token-based program reduction (DD-Token) gives the minimal program in \Cref{code:explanation_main}c, which suggests the argument \texttt{args} has an important role in the target prediction. However, with the AST-based program reduction (\perses), the minimal program is \Cref{code:explanation_main}d that highlights the signature of the method, for which \cts still can predict the same target label. Having these multiple explanations can improve the transparency of models inference.

\observation{Different program reduction techniques may provide additional explanations for better transparency of a specific prediction.}

\begin{table*} 
    \begin{center}
        \def\arraystretch{1.2}
        \caption{Adversarial evaluation with key features.}
        \label{table:key_adversarial}
        \begin{tabular}{|c|c|c|c|c|c|c|}
            \hline
             \multirow{2}{*}{\textbf{Reduction}} & \multirow{2}{*}{\textbf{Model}} & \multirow{2}{*}{\textbf{Adversarial Set}} & \multirow{2}{*}{\textbf{\#Initial}} & \multirow{2}{*}{\textbf{\#Transformed}} & \multicolumn{2}{c|}{\textbf{Misprediction}} \\ \cline{6-7}
             & & & & & \# & \% \\ \hline
            \hline
            
            \multirow{6}{*}{\ddd}
            & \multirow{3}{*}{\ctv} 
            & Actual Sample & 328 & 1148 & 135 & 11.76 \\ \cline{3-7}
            & & Key Token & 328 & 722 & 107 & 14.82 \\ \cline{3-7}
            & & Reduced Sample & 328 & 379 & 117 & 30.87 \\ \cline{2-7}
            & \multirow{3}{*}{\cts} 
            & Actual Sample & 287 & 836 & 109 & 13.04 \\ \cline{3-7}
            & & Key Token & 287 & 530 & 102 & 19.25 \\ \cline{3-7}
            & & Reduced Sample & 287 & 267 & 134 & 50.19 \\ \hline
            \hline
            
            \multirow{6}{*}{\perses}
            & \multirow{3}{*}{\ctv} 
            & Actual Sample & 320 & 911 & 97 & 10.65 \\ \cline{3-7}
            & & Key Token & 320 & 320 & 58 & 18.12 \\ \cline{3-7}
            & & Reduced Sample & 320 & 253 & 118 & 46.64 \\ \cline{2-7}
            & \multirow{3}{*}{\cts} 
            & Actual Sample & 280 & 658 & 93 & 14.13 \\ \cline{3-7}
            & & Key Token & 280 & 211 & 80 & 37.91 \\ \cline{3-7}
            & & Reduced Sample & 280 & 160 & 104 & 65.00 \\ \hline
        \end{tabular}%
    \end{center}
\end{table*}

\subsection{Key Targeted Adversarial Attacks on Models}
\label{sec:results_adverserial}

Here, we highlight the importance of key input features in programs by evaluating the adversarial generalizability \cite{rabin2021generalizability} or robustness \cite{yefet2020adversarial} of CI models in terms of the extracted key input features. We generate adversarial examples by applying semantic-preserving variable renaming transformation on programs, similar to \cite{rabin2021generalizability}, where we separately change each variable and all of its occurrences in the program with token \texttt{var}. We particularly compare the prediction of CI models before and after the variable renaming. In this experiment, we generate three types of adversarial sets: actual set, key set, and reduced set. First, in actual set, we target the actual initial programs and generate candidate transformed programs by considering all variables. Second, in key set, we also target the actual initial programs but generate candidate transformed programs by considering variables that occur in the key feature list. Third, in reduced set, we directly target the reduced programs for generating candidate transformed programs. 
The results of change in prediction (misprediction) for variable renaming transformation are shown in \Cref{table:key_adversarial}. 

According to \Cref{table:key_adversarial}, on average, the number of generated candidate transformed programs from the actual set are around $3x$ times higher than the initial programs, however, only $12\%$ of them trigger misprediction. Next, the number of generated candidate transformed programs from the key set are around $1.5x$ times higher than the initial programs and trigger $22\%$ misprediction. Although the key adversarial set contains $50\%$ less candidate transformed programs than the actual adversarial set, they trigger $10\%$ more misprediction. On the other hand, the reduced programs are the minimal program that CI models keep for preserving their target prediction. Therefore, the number of generated candidates transformed programs from the reduced set are the lowest as there are fewer tokens to apply transformations. However, the transformation on reduced programs is more powerful and triggers the highest percentage of misprediction. Moreover, comparing between \ddd and \perses, in most cases, \perses generated candidates transformed programs shows a higher rate of misprediction than \ddd. 

\observation{The adversarial programs based on key input features trigger $10\%$ more misprediction with $50\%$ fewer candidates. The \perses generated candidate programs are more vulnerable to adversarial transformation than \ddd, thus, highlighting the importance of key input features in programs.}

\section{Threats to Validity and Future Plan}
\label{sec:results_threats}

\Part{Evaluation}. We evaluated our approach for \mnp task with two CI models, four input types of randomly selected input programs, and Top-10 most frequent method names. Despite our best effort, it is possible that experiments with different models, tasks, and datasets may produce different results. Our further plan includes a detailed study with a variety of models, tasks, and larger datasets. 

\Part{Challenges}. 
One challenge we have for running \perses is that it loads the model in each reduction step while \ddd loads the model once at the beginning of reduction. For a fair comparison between them, we only consider the program reduction time and ignore the model loading time. We are working on optimizing the model loading time for \perses.
Another challenge for running \ddd, when there are multiple subsets that hold the same target criteria, \ddd sometimes gets stuck at that point. To keep the reduction process working, we temporarily used a timer to kill the current step and jump to the next step.

\Part{Artifacts}. We will publicly share the artifacts of this study at \href{https://github.com/mdrafiqulrabin/rm-dd-perses}{\color{blue}{https://github.com/mdrafiqulrabin/rm-dd-perses}}.

\section{Conclusion}
\label{sec:conclusion}

In this paper, we apply the syntax-guided program reduction technique, \perses, for reducing an input program while preserving the same prediction of the CI model.
The goal is to extract label-specific key input features of target labels for CI models from syntax-guided reduced programs.
We evaluate \perses on two popular CI models across four types of input programs for the method name prediction task.
Our results suggest that the syntax-guided program reduction technique (\perses) significantly outperforms the syntax-unaware program reduction technique (\ddd) in reducing different input programs.
Moreover, we extract key input features that CI models learn for a target label, by reducing some input programs of that label using \perses. The result shows that \perses mostly keeps label-specific key input features in its syntax-guided reduced programs than in \ddd's syntax-unaware reduced programs.
We also observe that the syntax-guided candidate programs are more vulnerable to adversarial transformation when renaming the key tokens in programs.
By identifying those key input features, we can better understand the learned behaviors of CI models from multiple explanations, which may improve the trustworthiness of models to correct prediction.

\section*{Acknowledgement}
This study has been done as coursework in the Department of Computer Science at the University of Houston (Course: COSC 6321 - Research Methods in Computer Science; Instructor: Omprakash D Gnawali). We organized an in-class conference (\href{https://sites.google.com/view/rq21}{\color{blue}{Research Quest 2021}}) and submitted our posters/papers as homework to the conference.

\balance
\bibliography{refs}
\bibliographystyle{IEEEtranN}

\end{document}